# 360º Phase Detector Cell for Measurement Systems based on Switched Dual Multipliers

Baltasar Pérez, Víctor A. Araña, Javier Perez-Mato, Francisco Cabrera

*Abstract—* This paper presents a 360º phase detector cell for performing phase-shift measurements on multiple output systems. An analog phase detector, capable of detecting a maximum range of ±90º, has been used to perform a double multiplication of two signals, both in-phase and phase-shifted. The proposed solution broadens the frequency range beyond other solutions that require to fulfil the quadrature condition. Subsequently, the possibility of reaching the theoretical limit of phase shift within a hybrid coupler (φ<90º±90º) is discussed by using four straight-line equations to characterise the phase detector response. The proposed solution allows to extend up to 360º the phase detection range and provide an increased immunity with respect to both impedance mismatching and phase deviations within the hybrid coupler. To demonstrate the feasibility of the proposed design, a phase detector cell prototype has been implemented using a commercial hybrid coupler with a phase shift of 92.5º±0.5º at 3.1 GHz to 5.9 GHz, an external switch and a microcontroller with 2KB of memory. Measurements show a range of detection of 360º (±180º) across the tested frequency band of 2.7 GHz to 6 GHz.

## I. INTRODUCTION

Measurement systems are characterised by their wide bandwidth and calibration capability. Phase shift measurements are widely used to evaluate or characterise those circuits where phase shifting across its elements becomes an essential characteristic of their behaviour, such as phase shifters, beamforming circuits, modulators, demodulators, and amplifiers. When the monitorization of multiple outputs is required, detectors may be grouped [1,2]. When the operating frequency increases, the possibility of using digital circuits capable of detecting phase shifts of 360º or more gets reduced. In these cases, the choices are limited to multiplier circuits that have a maximum phase shift detection range of ±90º. These analog phase detectors have been combined with 90º phase shifters in order to cover a detection range of 360º [3]. These circuits present a limitation on the operating bandwidth and therefore require a thorough adjustment of their internal voltages. Lastly, the use of down-converters, which can reduce the operating frequency to allow the use of digital detectors, would increase the complexity of the circuit and introduce isolation problems derived from the power levels required by the oscillator to correctly perform the signal mix.

In order to increase phase range detection using analog phase detectors and therefore avoid the frequency limitations of 90º phase shifters typically used in 360º phase detection, a new design based on the double multiplication of two signals in-phase and phase-shifted is proposed in this paper. Once calibrated, problems derived from impedance mismatching and 90º phase shift deviation can be supressed, making use of a simple bespoke control software. Due to the fact that multipliers and switches are common circuits, the solution is also valid for higher frequency bands. First, a discussion will be done regarding the problem associated with the use of multipliers as phase detectors on wideband systems, as well as problems derived from impedance mismatching. Additionally, the relation between measured phase error and the deviation from 90º in quadrature signal is evaluated. Finally, key details about the design are provided and a prototype cell will be constructed and evaluated to perform measurements on systems with multiple outputs.

## II. MISMATCHING AND FREQUENCY BAND LIMITATIONS IN 90º MULTIPLIER PHASE DETECTORS

To measure the phase shift ($\Delta\alpha$) between two high frequency signals with respect to the in-phase condition (0º), it's typical to use a 90º phase shifter, a multiplier and a low-pass filter (Fig. 1a). Furthermore, for the particular case of measuring the response of an array, the phase shifter on each cell would be substituted by a 90º hybrid in order to be able to combine the signals of adjacent locations [4,5]. The maximum detectable phase shift is given by the non-ambiguous zone within the ±90º range (Fig. 1b). By assuming $Vd=\sin(\Delta\alpha)$ and solving for $\Delta\alpha$, the phase shift can be calculated as $\Delta\alpha=\text{asin}(Vd)$, where $|\Delta\alpha|\leq 90º$.

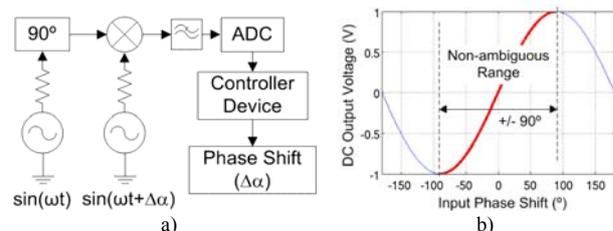

Fig. 1. a) Phase shift measurement circuit with phase shifter and multiplier. b) Ideal phase detector response highlighting the non-ambiguous range.

This work was supported by the Spanish Government under the TEC2011-29264-C03-02 and TEC2014-60283-C3-2-R project. All the authors are with IDeTIC-ULPGC, 35017, Canary Islands, Spain (e-mail: varana@idetic.eu).

However, frequency response of devices, impedance mismatching or phase deviation of hybrid couplers causes a reduction of this phase detection range. In this case, the non-ambiguous symmetric range around zero gets reduced by β as the phase shift increases with respect to the ideal case (Fig. 2), that is, Δα=asin(Vd), where |Δα|≤90-β.

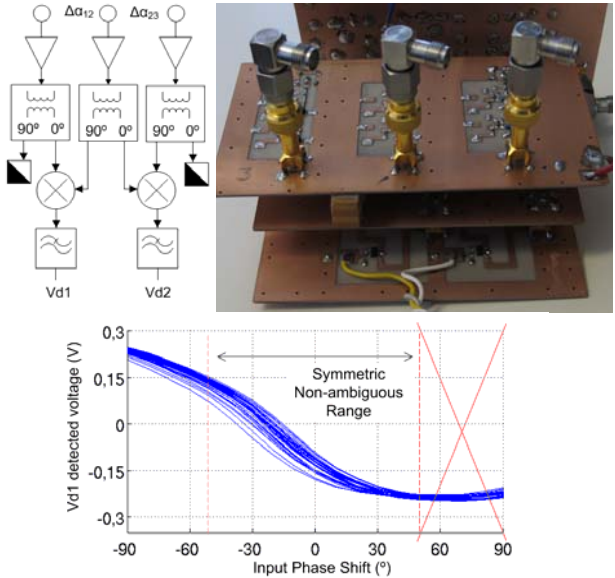

Fig. 2. Previous development based on Pogorzelski's work [4] (upper). Detector response vs input phase shift |Δα| ≤ 90º at several frequencies (4.85 GHz to 5.05 GHz). Symmetric detection range referred to in-phase condition (0º) is reduced by β=40º (lower).

### III. DUAL MULTIPLIER DESIGN THEORY

The proposed solution uses output voltages obtained by the multiplication of in-phase (I) and quadrature signals (Q) (Fig. 3) to overcome the ambiguity in phase shift measurements and exploit the performance offered by mixers in all frequency bands. Furthermore, the calculation of the phase shift is performed in the most linear area of each sinusoid in order to reduce noise-induced errors on the detected signal. Such a linear approach that minimizes the error for a given phase range (Linearized Range: LR) also reduces interpolation time and microcontroller complexity. By using this method a simple linear multi-section interpolation should be enough to compute the phase shift. The calculation of this phase shift requires a simple algorithm (Fig. 4) which selects the appropriate section (2 for each I/Q curves), without incurring a performance reduction for phase shift measurement systems.

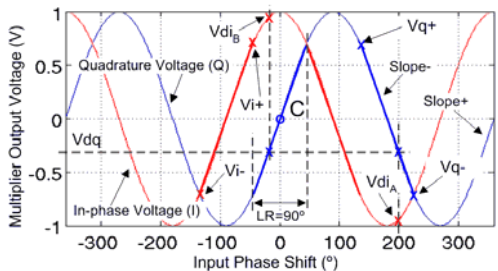

Fig. 3. Ideal response of I and Q multiplications. The area used to obtain the phase measurement by means of interpolation is highlighted. Zero voltage is considered for quadrature fulfillment (C) to simplify the explanation.

Regarding Fig. 3, for a given output quadrature voltage (Vdq) between quadrature LR (Vq+ and Vq-), there are two possible solutions given by Slope+ and Slope-. If the output in-phase voltage (Vdi) is around $Vdi_A$, the correct solution can be calculated from Slope- (200º). Otherwise, $Vdi_B$ leads to the Slope+ solution (-20º). A similar procedure can be followed with Vdi when this value has to be used to interpolate.

```
Read  Vdi and Vdq      %Normalized Output I and Q Voltage
If  abs(Vdq )≤ abs(Vdi)   Then     %Use Vdq to interpolate
    If  Vdi ≥ 0 Then    Interpolate in LR Quadrature Slope+
    Else  Interpolate in LR Quadrature Slope-    End If
Else    %Use Vdi to interpolate
    If  Vdq ≥ 0  Then  Interpolate in LR In-phase Slope-
    Else   Interpolate in LR In-phase Slope+      End If
End If
```

Fig. 4. Simplified algorithm to determine the input phase shift .

When sinusoids in Figure 3 are shifted by 90º, an LR equal to 90º is enough to cover 360º input phase shift. If not, it would be necessary to increase the LR. For example, in the case of a positive deviation of +30º from 90º, the necessary LR is 120º (Fig. 5). The light algorithm shown in Fig. 4 can be theoretically used for deviations of less than ±90º, although extreme values are not recommended due to voltage noise and non-ideal sinusoidal response in practical circuits. Furthermore, the linear approximation of the sinusoid implies that the error will increase with the LR. Therefore, new linear approximation with minimum error is done for each LR value.

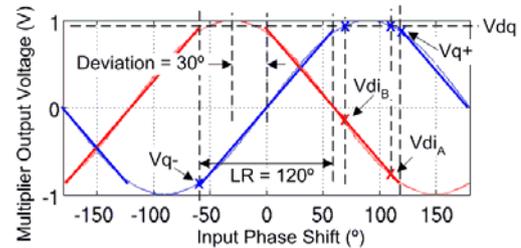

Fig. 5. I/Q curves for 30º deviation from the ideal 90º. Vdi=$Vdi_A$ leads to 110º solution and Vdi=$Vdi_B$ leads to 70º solution.

The minimum error (1.216º) is reached when the LR is 90º (±45º), which is the minimum LR value that would guarantee a 360º detection range. Fig. 6 shows the relation between both the maximum error and the phase deviation from 90º versus LR. As the LR increases, the maximum error increases, but the maximum admissible deviation can also increase. Depending on the application, the maximum admissible error and maximum associated quadrature deviation can be fixed as design target. For example, ±30º maximum deviation can be used if ±3º maximum error is accepted. In that case, an LR equal to 120º has to be used.

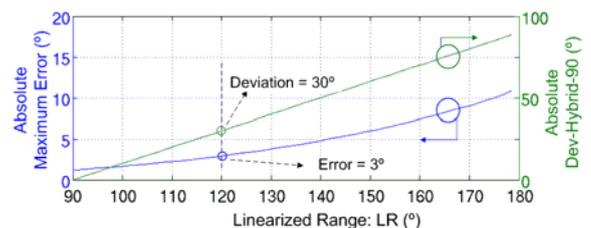

Fig. 6. Maximum error and phase deviation from 90º vs LR.

## IV. Experimental Results

The manufactured cell circuit which includes the two necessary branches to perform the required measurements is shown in Fig. 7. Each branch is composed of a 90º hybrid coupler (QCS-592 @92.5º±0.5º at 3.1-5.9 GHz), an RF switch to select 0º or 90º signal, and a 3 dB power splitter to deliver the signal to adjacent cells. The two adjacent outputs of each branch are connected to the multiplier detector inputs, followed by a signal conditioner (amp+DC shifter). Finally, a small microcontroller with 2KB memory is used to control the RF switches to obtain both the in-phase (Vdi with SW1 in 0º and SW0 in 0º) and quadrature-phase (Vdq with SW1 in 90º and SW0 in 0º) voltage using an internal 10-bit ADC and calculates the phase shift between the two input signals by using the algorithm described in Fig. 4. In order to test the system's goodness and robustness, one of the four capacitors at the output of the hybrid coupler has been changed from its initial design value of 10pF (balanced circuit) to 4.7pF (unbalanced circuit) to force a large phase deviation from 90º.

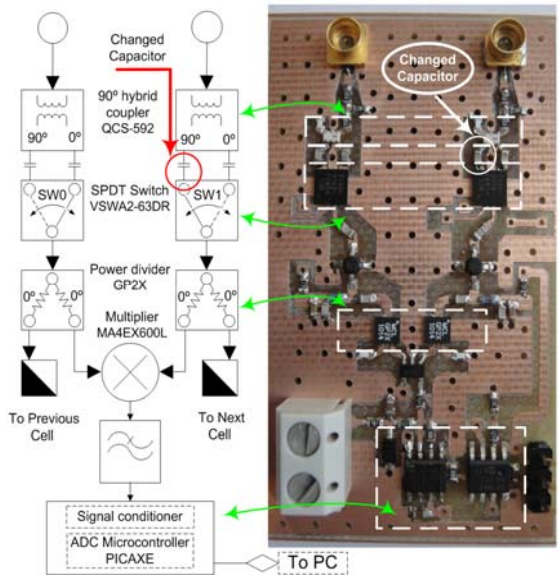

Fig. 7. Schematic and picture of the actual manufactured circuit based on switched dual multipliers design.

Previously, the proposed system needs to be calibrated in order to obtain the associated variables of the four linearized sections which allows to calculate the phase shift. The calibration procedure consists in inserting two calibration signals at the input ports, which are provided by two synchronised and calibrated vector signal generators [6]. As a first step, the I and Q voltages are captured while the generators' phase shift is swept between 0º to 360º. Afterwards, the phase deviation from 90º is computed, as well as the linear approaches, which minimize the error for the associated LR given by Fig. 6. Once the variables have been updated within the microcontroller, the behaviour of the developed prototype can be evaluated.

Figure 8 shows the measured phase error, as well as the Q output voltage and the I output voltage versus input phase shift with 10pF (5.75º deviation) and 4.7pF (38º deviation) at 5GHz. The theoretical error for a 38º deviation (Fig. 6) is ±3.795º and the measured maximum error is ±6.8º. The difference is due to the fact that the response of the detector is not a perfect sinusoid. In fact, the theoretical error for a 5.75º deviation is ± 1.46º and the measured maximum error is ±3.03º. Frequency response has also been obtained for balanced circuit (10pF), taking into account that specific calibration for each frequency is required. The most significant results are: ±2.25º error, 0.63º deviation at 2.7 GHz; ±1.3º error, 3.97º deviation at 6GHz; and ±5.72º error, 3.38º deviation at 4.1GHz that represents the worst case. The maximum tested frequency was limited by available signal generators and minimum frequency by device return losses.

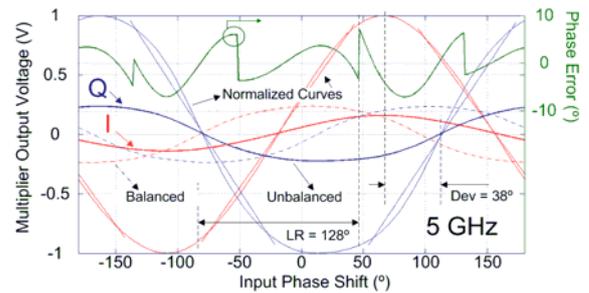

Fig. 8. Detected I/Q curves for balanced and unbalanced circuit. Phase error, normalized curves and linearized sections are showed for unbalanced circuit.

## V. Conclusion

Switched dual multipliers have been designed to increase analog phase detector range up to 360º. As quadrature condition fulfilment is not required, circuits in charge of generating I/Q curves can have a much broader operational frequency range. Moreover, the solution can be easily extended to higher frequency bands as it does not rely on digital circuits. A linearized characterization of output detector curves has been used at the expense of accuracy, in order to accelerate computation speed and reduce both interpolation time and microcontroller complexity. These characteristics make this solution especially useful for enabling fast control on antenna arrays, where the mean beam direction needs to be mostly maintained within a certain range.

## Acknowledgment

The authors would like to thank Juan D. Santana for his invaluable help during the manufacturing process.

## References

[1] P. Liao, R. A. York, "A Six-Element Beam-Scanning Array", *IEEE Microwave and Guided Wave Letters,* vol. 4, no. 1, pp. 20-22, Jan. 1994.
[2] L. Zhao, X. Gao, X. Hu, S. Liu, Q. An, "Beam Position and Phase Measurement System for Proton Accelerator in ADS", *IEEE Transactions on Nuclear Science*, vol. 61, no. 1, Feb. 2014.
[3] C. Tang, Q. Xue, "S-Band Full 360º High Precision Phase Detector", *Proceedings of APMC 2012, Kaohsiung, Taiwan*, pp. 97-99, Dec. 2012.
[4] R. J. Pogorzelski, "A Two-Dimensional Coupled Oscillator Array", *IEEE Microwave and Guided Wave Letters,* vol. 10, pp. 478-480, Nov. 2000.
[5] R. J. Pogorzelski, "A 5-by-5 Element Coupled Oscillator-Based Phased Array", *IEEE Trans. Antennas Propag.*, vol. 53, no. 4, pp. 1337-1345, Apr. 2005.
[6] P. Umpiérrez, V. Araña, F. Ramírez, "Experimental Characterization of Oscillator Circuits for Reduced-Order Models", *IEEE Trans. Microw. Theory Techn.*, vol. 60, no.11, pp. 3527-3541, Nov. 2012.